# LES STRUCTURES GÉOMÉTRIQUES ÉLÉMENTAIRES DE L'INFORMATION DIGITALE ÉLECTROMAGNÉTIQUE
*Caractérisation statistique de la mesure digitale des fluctuations spatio-Doppler et polarimétrique de l'onde électromagnétique radar*

*Frédéric Barbaresco[1], Yann Cabanes[1,2]*
[1]*Thales Land & Air Systems, Advanced Radar Concepts,* frederic.barbaresco@thalesgroup.com
[2]*Institut de Mathématiques de Bordeaux,* yann.cabanes@gmail.com

*Mots clés : Onde électromagnétique, Radar, Fluctuation Statistique, Géométrie de l'Information, classification non-supervisée, fouillis radar, Electromagnetic wave, Statistical Fluctuations, Information Geometry, unsupervised classification, radar clutter*

**Résumé/Abstract**

Il s'agit de décrire de nouvelles approches géométriques pour définir les statistiques de mesures spatio-temporelles et polarimétrique des états d'une onde électromagnétique, en utilisant les travaux de Maurice Fréchet, Jean-Louis Koszul et Jean-Marie Souriau, avec en particulier la notion d'état « moyen » de cette mesure digitale comme barycentre de Fréchet dans un espace métrique et un modèle issu de la mécanique statistique pour définir et calculer une densité à maximum d'entropie (extension de la notion de gaussienne) pour décrire les fluctuations de l'onde électromagnétique. L'article illustrera ces outils nouveaux avec des exemples d'application en radar pour la mesure Doppler, spatio-temporelle et polarimétrique de l'onde électromagnétique en introduisant une distance sur les matrices de covariance du signal digital électromagnétique, basé sur la métrique de Fisher issue de la Géométrie de l'Information.

## 1   La mesure digitale de l'onde électromagnétique radar

Au XXième siècle, l'Homme a domestiqué les ondes électromagnétiques qui ont profondément bouleversé ses modes de vies pour communiquer, localiser et observer la Nature et l'Univers. Dans ce contexte, le RADAR a une longue histoire plus que centenaire mais la **digitalisation récente spatiale et temporelle complète**, de la mesure des ondes électromagnétiques par **des antennes tout numériques** ouvre la voie à la manipulation plus efficace de l'Information inscrite dans ce signal aléatoire. Cette digitalisation spatio-temporelle de l'onde électromagnétique a révélé les structures géométriques intimes qui décrivent cette Information. Ces structures révèlent des affinités naturelles avec les outils mathématiques les plus récents comme la géométrie des espaces métriques et la géométrie Symplectique. Nous parcourrons ces avancées récentes du traitement du signal électromagnétique RADAR. Nous soulignerons au fil de l'exposé comment interviennent les intuitions lumineuses et le plus souvent oubliées présentes dans cette filiation des idées de Clairaut, Legendre, Massieu, Poincaré, Cartan, Fréchet, Koszul, Balian et Souriau, pour décrire les outils géométriques développés pour caractériser les fluctuations statistiques de l'onde électromagnétique, et permettant de décrire des densités de probabilités pour ces états aléatoires localement stationnaires de l'onde.

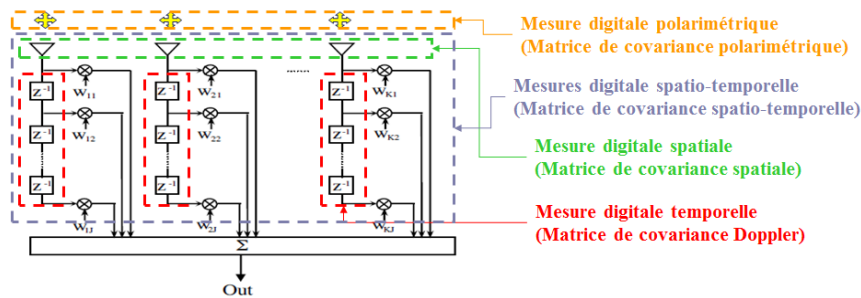

*Figure 1 : illustration d'antenne tout numérique :
digitalisation spatiale, temporelle et polarimétrique de la mesure de l'onde électromagnétique*

## 2   Caractérisation statistique de la mesure digitale des fluctuations de l'onde électromagnétique radar

L'idée, qui sera exposé dans l'article, est de « coder » l'état stationnaire d'une mesure digitale de l'onde électromagnétique (Direction, Doppler et Polarimétrie) par un point dans un espace métrique. Comme il s'agit d'une mesure statistique, nous montrerons que la métrique la plus « naturelle » est donnée en Géométrie de l'Information par la métrique de Fisher du processus stationnaire.

L'extension des statistiques dans les espaces métriques a été introduite par Maurice Fréchet à travers la notion « d'espace distancié ». L'approche par les espaces métriques permet de définir la moyenne ou la médiane d'éléments dans un espace distancié par le barycentre géodésique de Fréchet calculé par le flot de gradient introduit dans les années 70 par Hermann Karcher. Pour définir ces espaces distanciés pour une onde électromagnétique, il nous faut au préalable introduire une distance entre des mesures digitales des états de l'onde. La métrique de Fisher de la « Géométrie de l'Information » fournit une distance naturelle, disposant des qualités d'invariances souhaitées (invariance par reparamétrisation, invariance liée aux symétries du signal électromagnétique). La métrique de Fisher peut être généralisée dans les espaces décrivant les états de l'onde électromagnétique à partir des structures géométriques introduites par Jean-Louis Koszul (fonction caractéristique de Koszul-Vinberg, 2-forme de Koszul). L'étape suivante consiste à généraliser la notion de densité gaussienne pour les états de l'onde électromagnétiques, en généralisant la notion de densités à maximum d'entropie (densité de Gibbs), ce qui nous oblige à généraliser la définition de l'entropie. Partant des fonctions caractéristiques de François Massieu (introduite en thermodynamiques par François Massieu), cette généralisation a été établie par Jean-Marie Souriau en Mécanique Statistique dans le cadre de la Mécanique Géométrique, avec un modèle qui a été appelé « thermodynamique des groupes de Lie », dont on déduit une densité (de Gibbs) de probabilités et une métrique de Fisher-Souriau pour décrire l'état moyen (barycentre de Fréchet) des fluctuations spatio-Doppler et polarimétrique des mesures digitales de l'onde électromagnétique.

Dans le cas d'un signal stationnaire, en utilisant, les structures Toeplitz (respectivement Blocs-Toeplitz) des matrices de covariance du signal temporel [Doppler] ou spatial [direction] (respectivement spatio-temporelle), on démontre que la métrique naturelle est Kählérienne dans l'espace produit du polydisque de Poincaré (codage de information Doppler) et du polydisque de Siegel (codage de l'information spatio-temporelle).

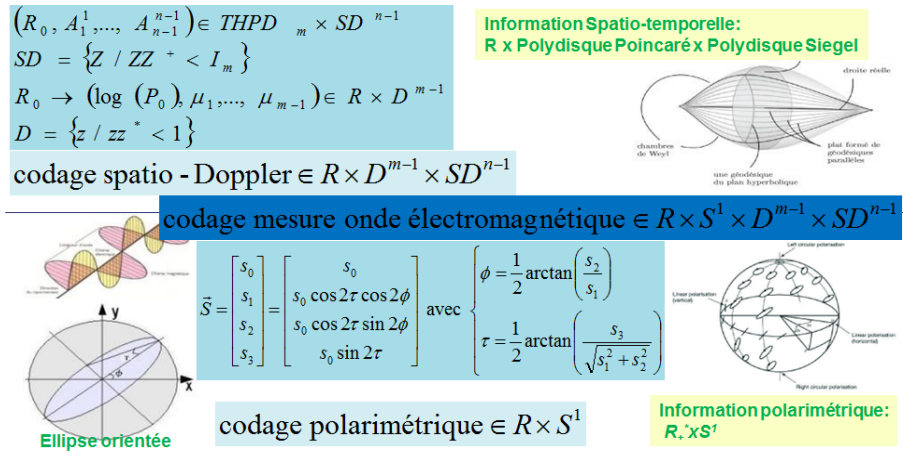

*Figure 2 : Codage de la mesure spatio-Doppler et polarimétrique dans l'espace métrique $\Re \times S^1 \times D^{m-1} \times SD^{n-1}$*

Dans le cas d'un signal localement stationnaire, on caractérise le signal spatio-temporel électromagnétique par un modèle autorégressif matriciel (extension matricielle du théorème de Trench-Verblunsky), basé sur la théorie MOPUC (Matrix Orthogonal Polynomial on the Unit Circle). A partir de la mesure digitale spatio-temporelle de l'onde électromagnétique, on estime une série de N matrices de covariance THDP (Toeplitz Hermitienne Définie Positive) $\{R_0,...R_{N-1}\}$. Cette série temporelle est ensuite paramétrée par un modèle autorégressif matriciel $\left(R_0, \{A_k^k\}_{k=1}^{N-1}\right) \in THPD \times SD^{N-1}$ avec $R_0 \in THPD_n$ une matrice THDP, et $A_k^k \in SD$, $k=1,...,N-1$ les coefficients de réflexion/Verblunsky matriciels dans le disque de Siegel unité $SD$, tel que $A_k^k \left(A_k^k\right)^+ < I$ (où + signifie trans-conjugué et I la matrice identité). La matrice $R_0$ peut elle-même être décrite par un modèle autorégressif classique complexe, de paramètres $\left(P_0, \{\mu_k\}_{k=1}^{m-1}\right) \in \Re^{+*} \times D^{m-1}$ avec $P_0$ le terme de puissance, dont on prend classiquement le logarithme $\log(P_0) \in \Re$ et $\mu_k \in D, k=1,...,m-1$ les coefficients de réflexion/Verblunsky dans le disque unité de Poincaré $D$, tel que $\mu_k \mu_k^* = |\mu_k|^2 < 1$. On a ainsi pour une onde électromagnétique localement stationnaire, un codage de l'information dans un espace métrique $\left(\log(P_0), \{\mu_k\}_{k=1}^{m-1}, \{A_k^k\}_{k=1}^{N-1}\right) \in \Re \times D^{m-1} \times SD^{n-1}$, à laquelle on peut rajouter $\mathbb{S}^2$ : la sphère de Poincaré pour l'information polarimétrique. Le signal étant finalement codé sur l'espace produit $\mathbb{R} \times \mathbb{S}^2 \times D^{m-1} \times SD^{n-1}$.

La théorie de la « Géométrie de l'Information » introduite par Rao et Fréchet permet de définir une métrique naturelle entre densité de probabilité dans l'espace des paramètres et qui est invariante par à rapport à un changement de paramétrisation. En particulier, on peut définir la métrique par le hessien de l'Entropie. En considérant le processus stationnaire décrivant l'état spatio-temporel de l'onde électromagnétique par la matrice de covariance du vecteur spatio-temporel et sa matrice de covariance associée Toeplitz-Blocs-Toeplitz :

$$R_{N,p} = \begin{bmatrix} R_0 & R_1 & \cdots & R_{N-1} \\ R_1^+ & R_0 & \ddots & \vdots \\ \vdots & \ddots & \ddots & R_1 \\ R_{N-1}^+ & \cdots & R_1^+ & R_0 \end{bmatrix}$$

l'entropie est donnée par $S(R_{p,N}) = \tilde{\Phi}(R_{p,N}) = -\log(\det R_{p,N}) + cste$, qui avec les paramètres autorégressifs matriciels s'écrit: $S(\bar{R}_{p,N}) = -\sum_{k=1}^{N-1}(N-k) \cdot \log \det\left[1 - \bar{A}_k^k \bar{A}_k^{k+}\right] - N \cdot \log\left[\pi \cdot e \cdot \det \bar{R}_0\right]$ où $\bar{X}$ représente l'estimation de $X$. En considérant, la paramétrisation $\theta^{(N-1)} = \begin{bmatrix} \bar{R}_0 & \bar{A}_1^1 & \cdots & \bar{A}_{N-1}^{N-1} \end{bmatrix}^T$, la métrique est alors donnée par le hessien de l'Entropie ce qui donne :

$$ds^2 = d\theta^{(N-1)+} \left[\frac{\partial^2 S}{\partial \theta_i^{(N-1)} \partial \theta_j^{(N-1)}}\right]_{i,j} d\theta^{(N-1)} = N \cdot Tr\left[\left(\bar{R}_0^{-1} d\bar{R}_0\right)^2\right] + \sum_{k=1}^{N-1}(N-k)Tr\left[\left(I_n - \bar{A}_k^k \bar{A}_k^{k+}\right)^{-1} d\bar{A}_k^k \left(I_n - \bar{A}_k^{k+} \bar{A}_k^k\right)^{-1} d\bar{A}_k^{k+}\right]$$

On peut intégrer cette métrique pour définir une distance en paramétrisation $\{\bar{R}_0, \bar{A}_1^1, ..., \bar{A}_{N-1}^{N-1}\}$ permettant de déterminer une distance entre 2 matrices de covariance spatio-temporelle de l'onde électromagnétique $R_{(1)}$ et $R_{(2)}$ :

$$\begin{cases} d^2(R_{(1)}, R_{(2)}) = N \cdot \left\|\log\left(\bar{R}_{(1),0}^{-1/2} \bar{R}_{(2),0} \bar{R}_{(1),0}^{-1/2}\right)\right\|_F^2 + \sum_{k=1}^{N-1}(N-k)\log^2\left(\frac{1 + \left\|\Phi_{A_{(1),k}^k}\left(\bar{A}_{(2),k}^k\right)\right\|}{1 - \left\|\Phi_{A_{(1),k}^k}\left(\bar{A}_{(2),k}^k\right)\right\|}\right) \\ \Phi_Z(W) = \left(I - ZZ^+\right)^{-1/2}(W - Z)\left(I - Z^+W\right)^{-1}\left(I - Z^+Z\right)^{1/2} \end{cases}$$

Dans l'article final, on décrira le modèle de Souriau pour définir une densité à Maximum d'Entropie pour les mesures d'états spatio-temporel de l'onde électromagnétique radar, qui est covariante sous l'action du groupe de Lie qui agit de façon homogène dans l'espace de représentation de l'onde électromagnétique.

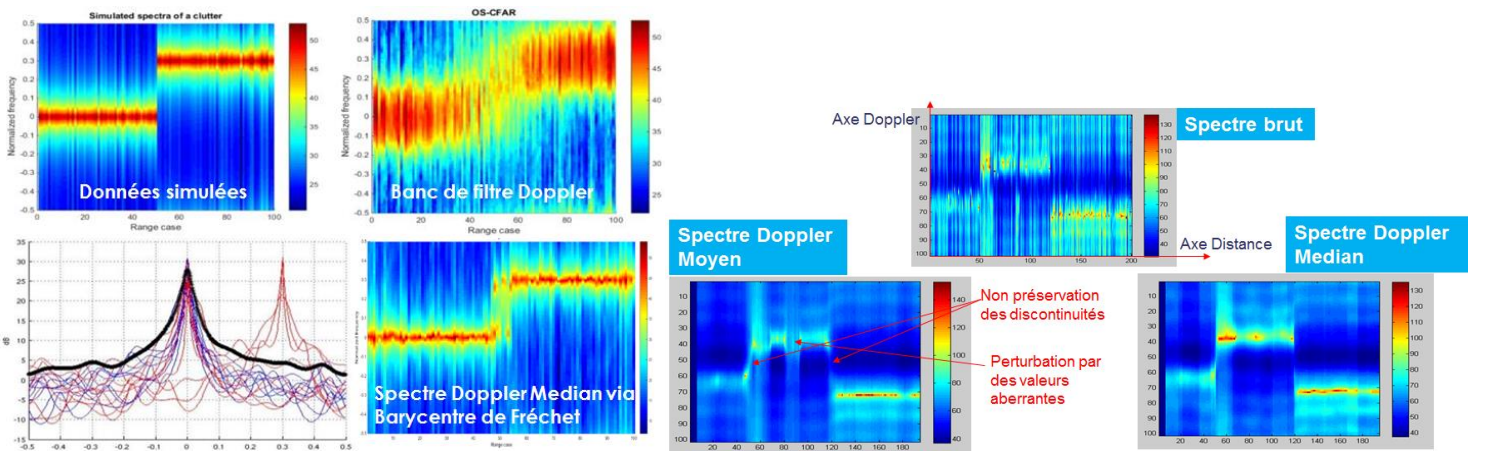

*Figure 3 : Exemple de calcul de spectre Doppler **médian** (défini comme barycentre géodésique de Fréchet via la métrique de Fisher) pour les matrices de covariance temporelle du signal Doppler de l'onde électromagnétique (on remarque la propriété de robustesse du « médian » à des valeurs parasites suivants l'axe distance)*

# 3 Application de l'analyse Doppler dans les espaces métriques à la classification non-supervisée du fouillis radar

## 3.1 Des données radar aux matrices complexes

Pour simplifier notre étude, les données d'entrée seront prises sur une seule rafale pour le faisceau d'élévation nulle. Le radar nous fournit donc une matrice complexe U de taille : (#impulsions) × (#cases distance).

Le nombre d'impulsion d'une rafale varie pour nos données entre 8 et 20 ; le nombre de cases distance varie également, il est de l'ordre de 1000. Une case distance mesure environ 60 mètres de longueur.

Le coefficient complexe $U_{i,j}$ représente l'amplitude et la phase après compression d'impulsion de l'écho revenant de la case distance i après l'impulsion j.

Les données que nous allons classifier sont les cases distance, chaque case distance étant représentée par une colonne de la matrice U.

## 3.2 Hypothèses de modélisation

Nous nous concentrons à présent sur une case distance, dont nous analysons le signal comme étant une série temporelle indexée par le numéro de l'impulsion.

Nous supposons que le signal est stationnaire dans la rafale et de moyenne nulle.

Nous supposons également que le signal peut être modélisé par un processus autorégressif Gaussien.

## 3.3 Espace métrique de représentation des données

Nous cherchons à classifier les cases distance en associant à chacune d'entre elle la matrice d'autocorrélation de la série temporelle associée, puis en classifiant les matrices d'autocorrélation.

Cela nous amène à construire un nouvel espace métrique dans lequel les matrices d'autocorrélation « proches » seront regroupées dans un même cluster.

Chaque matrice d'autocorrélation peut être représentée façon équivalente par des coefficients dans l'espace $\mathbb{R}_+^* \times D^{n-1}$, où $D$ est le disque unité complexe et n est le nombre d'impulsions.

Ces coefficients sont les coefficients d'un modèle autorégressif temporel ; ils sont calculés en utilisant l'algorithme de Burg régularisé.

Nous munissons cet espace d'une métrique naturelle issue de la géométrie de l'information.

## 3.4 Classification

### 3.4.1 *Simulation de données*

Pour pouvoir tester la performance des algorithmes de classification mis au point, il faut disposer de plusieurs jeux de données, chaque jeu étant homogène (réalisation de variables aléatoires iid). L'homogénéité est quelque chose de difficilement contrôlable concernant les données de terrain. Pour pallier à cette difficulté, nous avons mis au point un algorithme de simulation de données radar, en utilisant un modèle SIRV (Spherically Invariant Random Vectors). Ce modèle est paramétrique. Les paramètres utilisés pour simuler des données dépendent du type de fouillis que l'on souhaite simuler.

Pour tester les algorithmes de classification, nous simulons donc deux groupes de données, un premier groupe avec un paramètre P1, puis un second avec un paramètre P2. Un bon algorithme de classification regroupera dans un même cluster les données issues de P1, et dans un autre cluster les données issues de P2.

3.4.2 *Classification par l'agorithme des k-means*

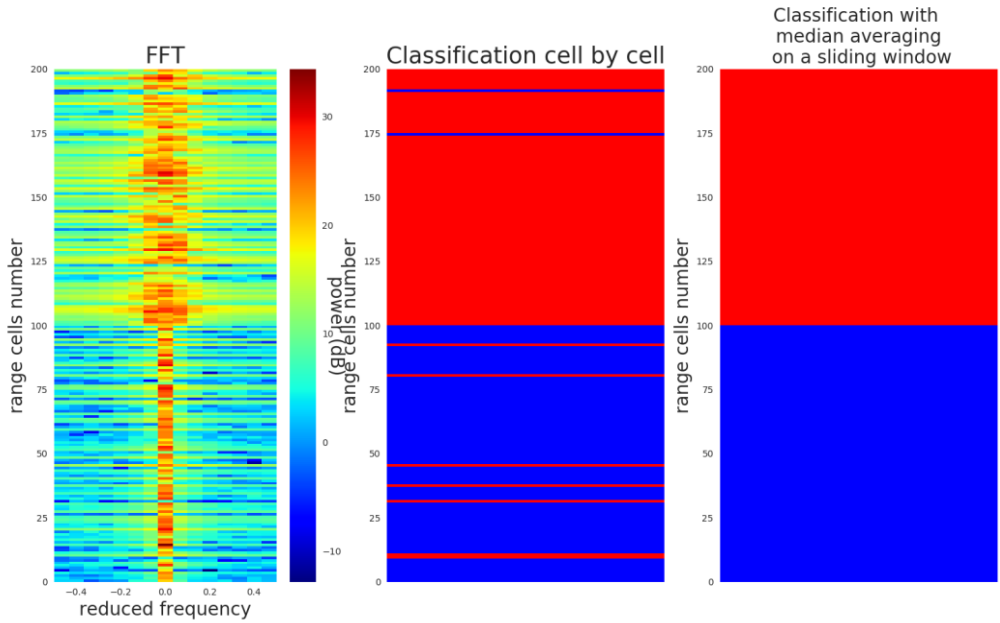

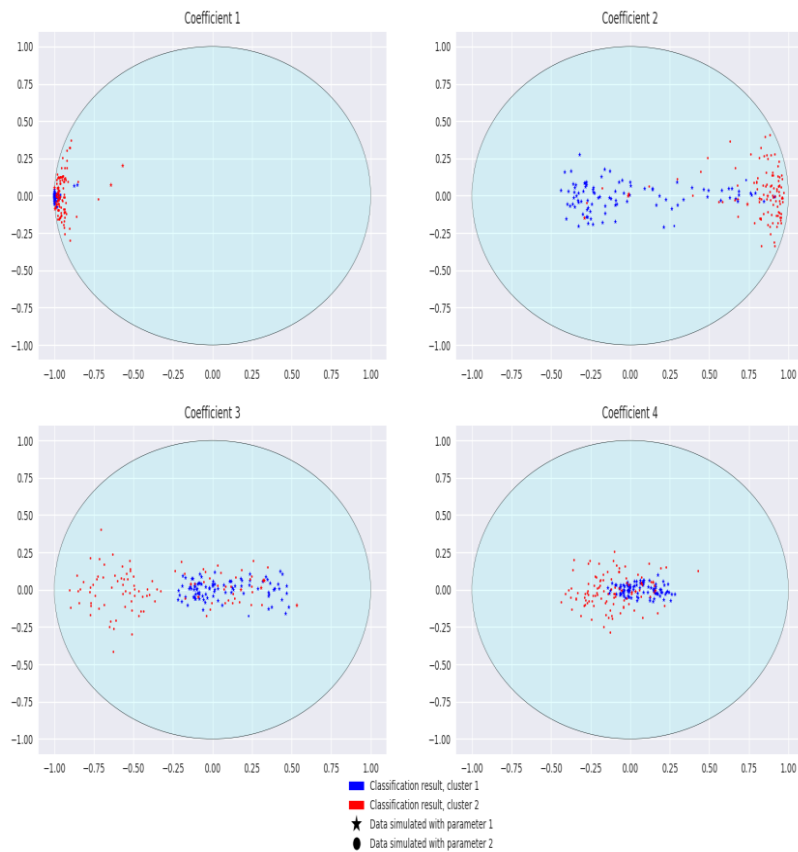

Nous fixons le nombre de clusters k souhaités au début de l'algorithme. Nous initialisons l'algorithme en choisissant k points de façon aléatoire parmi l'ensemble des points à classifié. Ils représentent maintenant des barycentres de classes.

A chaque itération de l'algorithme :

- Associer à chaque point au cluster dont le barycentre est le plus proche. Pour cette étape, il faut disposer d'une distance explicite sur notre espace.

- Calculer le nouveau barycentre du cluster. Pour cette étape, il faut pouvoir approximer la position du barycentre d'un nuage de points, celui-ci étant défini comme le point X qui minimise la somme des distances au carré entre X et les points du nuage. Ce calcul n'est pas évident dans les espaces métriques non-euclidiens.

3.4.3 *Méthode de visualisation de la performance des algorithmes de classification*

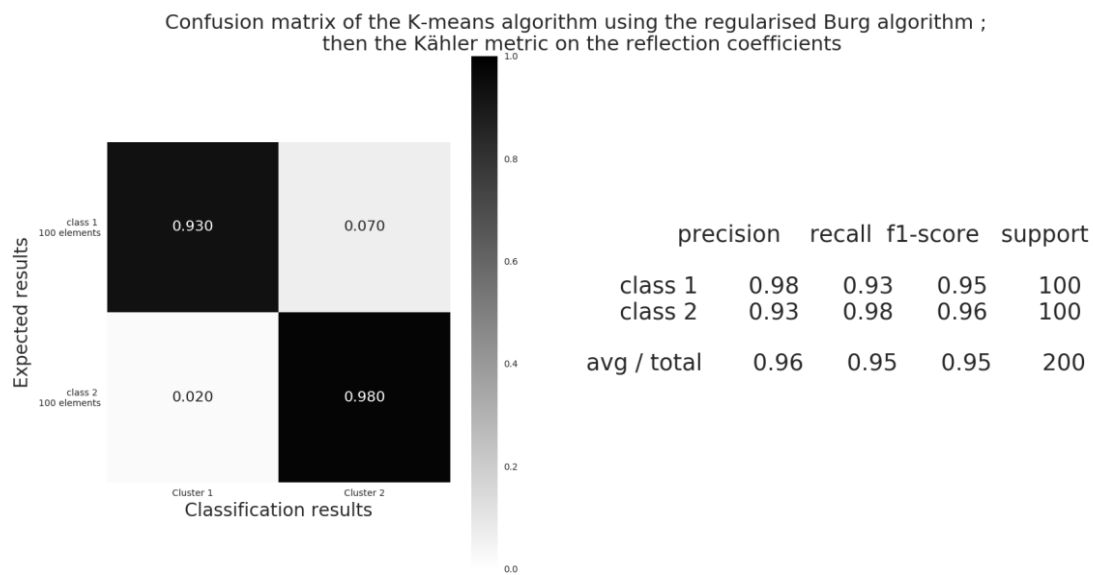

Pour visualiser et pouvoir comparer la performance des algorithmes de classification non-supervisée, nous avons emprunté les méthodes usuelles de visualisation des algorithmes de classification supervisés : la matrice de confusion pour la visualisation et le f1-score (calculé à partir de la précision et du recall) pour le score. Puisque notre classification non supervisée ne donne pas de labels comme « paramètre de simulation P1 » ou « paramètre de simulation P2 » à nos clusters, nous testons toutes les permutions possibles pour trouver quel cluster correspond à quel jeu de données simulées en utilisant le paramètre Pi. Le score de la classification correspond au score de la meilleure permutation.

**Références bibliographiques**